\documentclass[a4paper,10pt]{article}
\usepackage[hmargin=2cm,vmargin=1.6cm]{geometry}
\usepackage{booktabs}
\usepackage{amsmath, amssymb}

\usepackage{amsfonts}
\usepackage{float}
\usepackage[small,it]{caption}
\usepackage{multirow}
\usepackage{setspace}
\usepackage{pdfpages}
\usepackage{exscale,relsize}
\usepackage{subfig}
\usepackage{setspace}
\usepackage{empheq}
\usepackage[numbers,sort]{natbib}
\usepackage{comment}
\usepackage{tcolorbox}
\usepackage{bigints}
\usepackage{bm}
\usepackage{graphicx}
\usepackage{epstopdf}

\usepackage[framemethod=TikZ]{mdframed}
\usepackage{amsthm}
\newcounter{theo}[section] 
\renewcommand{\thetheo}{\arabic{section}.\arabic{theo}}

\makeatletter
\newsavebox\myboxA
\newsavebox\myboxB
\newlength\mylenA

\newcommand*\xoverline[2][0.75]{%
    \sbox{\myboxA}{$\m@th#2$}%
    \setbox\myboxB\null% Phantom box
    \ht\myboxB=\ht\myboxA%
    \dp\myboxB=\dp\myboxA%
    \wd\myboxB=#1\wd\myboxA% Scale phantom
    \sbox\myboxB{$\m@th\overline{\copy\myboxB}$}%  Overlined phantom
    \setlength\mylenA{\the\wd\myboxA}%   calc width diff
    \addtolength\mylenA{-\the\wd\myboxB}%
    \ifdim\wd\myboxB<\wd\myboxA%
       \rlap{\hskip 0.5\mylenA\usebox\myboxB}{\usebox\myboxA}%
    \else
        \hskip -0.5\mylenA\rlap{\usebox\myboxA}{\hskip 0.5\mylenA\usebox\myboxB}%
    \fi}
\makeatother

\newcount\colveccount
\newcommand*\colvec[1]{
        \global\colveccount#1
        \begin{pmatrix}
        \colvecnext
}
\def\colvecnext#1{
        #1
        \global\advance\colveccount-1
        \ifnum\colveccount>0
                \\
                \expandafter\colvecnext
        \else
                \end{pmatrix}
        \fi
}

\begin{document}

\title{Modelling the evolution of traits in a two-sex population, with an application to grandmothering}
\author{Matthew H.\ Chan$^{1}$,
\and Kristen Hawkes$^{2}$,
\and Peter S.\ Kim$^{3}$}
\date{\today}
\maketitle

\footnotetext[1]{School of Mathematics and Statistics, University of Sydney,
 NSW 2006, Australia. {\tt M.Chan@maths.usyd.edu.au}}
 \footnotetext[2]{Department of Anthropology, University of Utah, Salt Lake City,
 UT 84112, USA. {\tt Hawkes@anthro.utah.edu}}
   \footnotetext[3]{School of Mathematics and Statistics, University of Sydney,
 NSW 2006, Australia. {\tt PKim@maths.usyd.edu.au}}
\renewcommand{\thefootnote}{\arabic{footnote}}
\newenvironment{acknowledgements}{{\flushleft \bf Acknowledgements:}}{}

\begin{abstract}
We present a mathematical simplification for the evolutionary dynamics of a heritable trait within a two-sex population. This trait is assumed to control the timing of sex-specific life-history events, such as the age of sexual maturity and end of female fertility, and each sex has a distinct fitness tradeoff associated with the trait. We provide a formula for the fitness landscape of the population and show a natural extension of the result to an arbitrary number of heritable traits.  Our method can be viewed as a dynamical systems generalisation of the Price Equation to include two sexes{, age structure} and multiple traits. We use this formula to examine the effect of grandmothering, whereby post-fertile females subsidize their daughter's fertility by provisioning grandchildren. Grandmothering can drive a shift towards higher sex ratios due to lengthening female post-fertile longevity, leading to changes in fitness for both sexes. For males, increased longevity is accompanied by a substantially longer fertile lifespan resulting in higher sex ratios in the fertile ages. Our fitness landscapes show a net increase in fitness for both males and females at longer lifespans, and as a result, we find that grandmothering alone provides an evolutionary trajectory to higher longevities.
\end{abstract}


\section{Introduction}

The modelling of a two-sex population with heritable traits has often proved to be a formidable problem due to the interplay of many seemingly crucial evolutionary mechanisms. Our study is motivated by a specific example of this problem: the evolution of longevity in ancestral populations, where sex-specific life histories introduce biases in the mating sex ratio and each sex has a fitness tradeoff associated with longevity. The tradeoffs and life histories vary with longevity, which is expected from Charnov's model of {life history variation in female mammals} \cite{129}, and {when models include both males and females,} result in a \emph{sexual conflict}, wherein each sex is optimised at different longevities, but both optima cannot be obtained simultaneously \cite{173,174}. This leads to the selected longevity of the population being a compromise between the two optima. Although an observation shared by previous studies is that the sexual conflict plays a significant role in the evolutionary dynamics, it remains a major obstacle in gaining a deeper understanding of the problem as it has been difficult to explicitly quantify \cite{114,115,116,117,170}. This further adds to the confusion on exactly which components of the problem generate the sexual conflict and how strongly such components contribute to the sexual conflict.
\\
\\
Instead of reasoning about this problem in a discrete sense, for example with an agent-based approach, we show that it is possible, and more convenient, to consider continuous distributions along trait-space. This approach allows for a simplified computation of the evolutionary dynamics and the explicit calculation of the fitness landscape of the population. Moreover, the male and female fitnesses across trait-space can be separately calculated, which provides a way to visualise the sexual conflict. Since the calculation of the fitness landscape is explicit and inexpensive, comparisons of fitness landscapes generated from different sex-specific life-history strategies and tradeoffs can be made. In Section \ref{ntrait}, we generalise this result for an arbitrary number of heritable traits that affect the sex-specific tradeoffs and life-history parameters. {A common approach to such problems is discrete formulations aiming to generalise Price's equation (for examples see Batty et al. \cite{179} and Grafen, A. \cite{180}), which is a mathematical statement of how trait frequencies in a population change over time, given the initial population frequencies over trait space and the fitnesses of the traits (see Frank \cite{176} for a detailed discussion and also Price's original papers \cite{177,178}). We take an alternative dynamical-systems approach; thus, our result can be thought of as a dynamical-systems generalisation of Price's equation to include two sexes, age structure and multiple traits.}
\\
\\
In Section \ref{GMsect}, we apply the result to determine whether grandmothering, whereby post-fertile females increase the fertility rate of their daughters by provisioning their grandchildren, can drive the evolution of increased longevity by allowing for a transition between equilibria in the trait-space. This question stems from the Grandmother Hypothesis, which proposes increased longevity was selected for when ancestral populations began relying on foods weaned juveniles could not acquire effectively for themselves. Under such circumstances, mothers would have to feed their offspring longer, but with subsidies from grandmothers, they could have next offspring sooner; and longer-lived grandmothers, without infants themselves, could support more grandchildren.
\\
\\
Whether grandmothering alone can propel a great ape-like population to higher longevities has been the subject of several modelling studies recently, many of which involve computationally expensive simulations \cite{115,116,117,118}. Similar to Kim et al. \cite{116}, Kim et al. \cite{117} and Chan et al. \cite{170}, who considered the problem with comparable assumptions, we find here that grandmothering alone can drive the evolution of increased longevity. The key difference from previous models is that our approach produces an explicit fitness landscape which is inexpensive to compute. This allows for a straightforward comparison between {fitness landscapes of both sexes and the population compromise that evolves} with and without grandmothering.

\section{Model}
\subsection{Problem setup}
\label{setup}

We consider a two-sex population in which each individual possesses a heritable longevity trait value $x$. {Females are fertile from $\tau_f(x)$ to $\hat{\tau}_f(x)$, while males are fertile from $\tau_m(x)$ to $\hat{\tau}_m(x)$, where $\hat{\tau}_f(x)$ and $\hat{\tau}_m(x)$ denote the end of fertility for females and males respectively for a particular longevity trait value $x$. We make no assumptions on the form of $\tau_f(x)$, $\tau_m(x)$, $\hat{\tau}_f(x)$ or $\hat{\tau}_m(x)$, only that they are piecewise continuous functions.} Both males and females have equal age-specific mortality rates $\mu(x)$, leading to equivalent age-profiles for both sexes.
\\
\\
In each time interval, fertile males compete over every fertile female for a chance at paternity. The competitive success rate for a male with a specific trait value $x$ is governed by the male longevity-fertility tradeoff function $\phi(x)$, representing the relative probability that a male will out-compete others for a chance at paternity. Likewise, the female birth rate at each time $t$ is governed by the female longevity-fertility tradeoff function $b(x)$, {which accounts for varying interbirth intervals as longevity increases}. Offspring inherit the mean value of their parents' longevity trait values, with a probability to mutate according to a normal distribution with mean 0 and variance $\varepsilon^2$. We assume that there is equal probability for an offspring to be male or female, leading to an equal sex ratio. Finally, to ensure that the population converges to a finite equilibrium, we assume that the number of offspring with a specific trait value $x$ entering the population is regulated by a competition factor equal to the total number of births at time $t$.

\subsection{Model formulation}

\begin{figure}[H]
  \hspace{5mm}
  \centerline{
  \subfloat[]{\label{fig1}\includegraphics[width=0.55\textwidth]{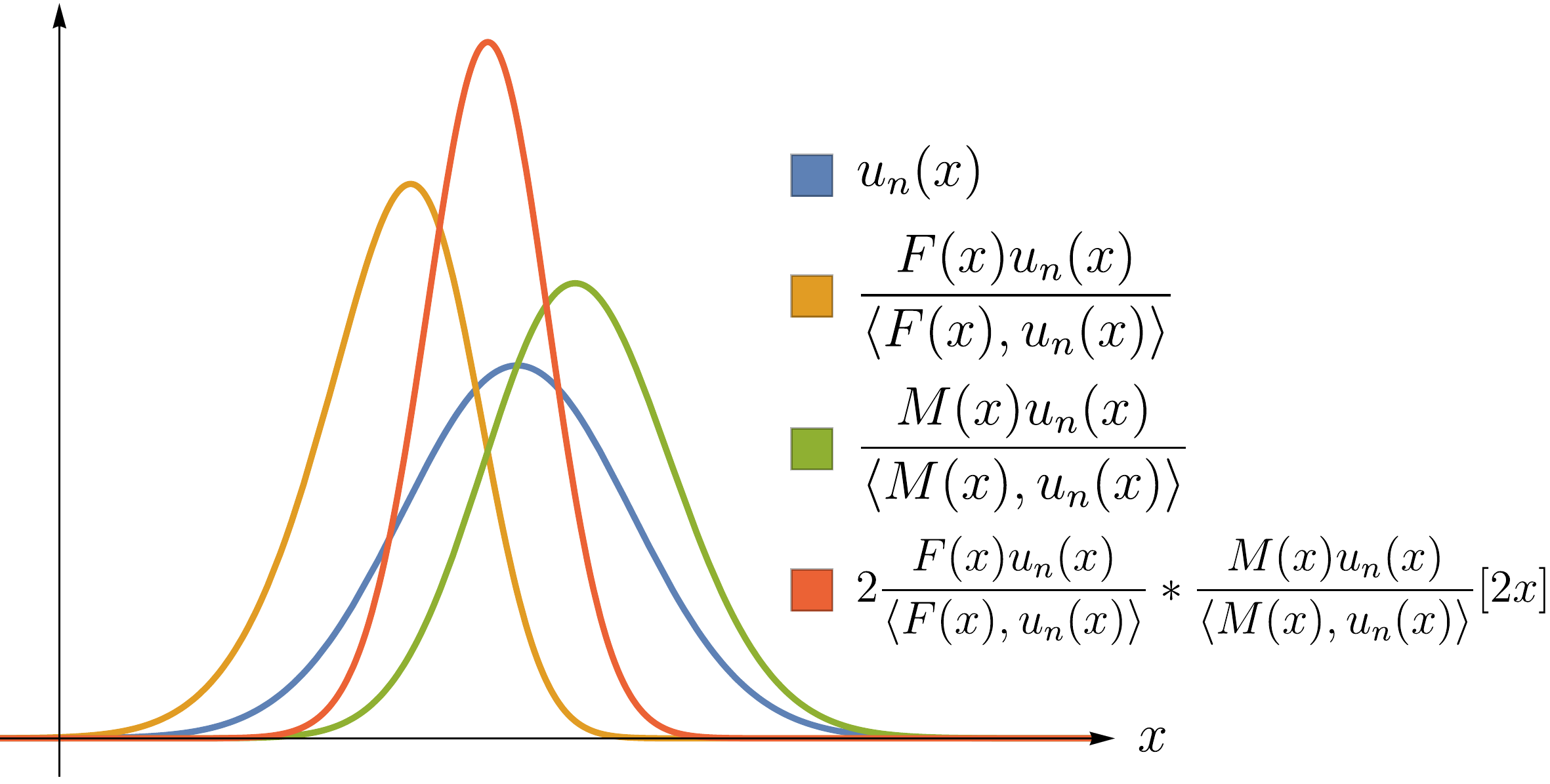}}
  \hspace{-0mm}
  \subfloat[]{\label{fig2}\includegraphics[width=0.55\textwidth]{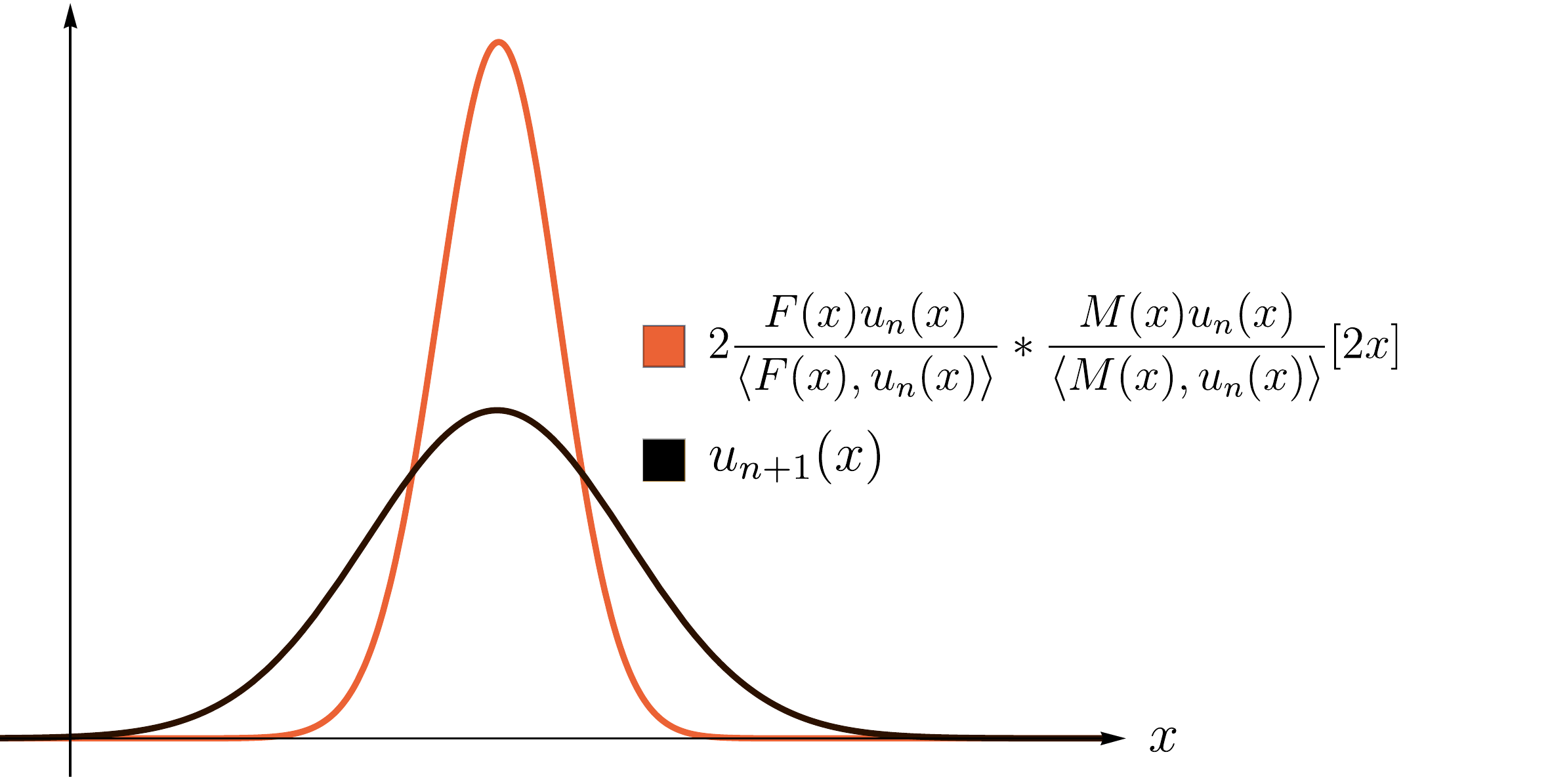}}}
  \caption{An illustration of the process described in Eq.~(\ref{convint}) for hypothetical functions $F(x)$ and $M(x)$. We note that the height of the individual curves is irrelevant; the important aspect is that they integrate to 1, since they describe distributions across trait-space. The black curve, representing $u_{n+1}(x)$, is obtained by convolving the function given in red with a normal density function with zero mean and finite variance.}
  \label{fig3}
\end{figure}

\noindent We let $u(a,x,t)$ denote the density of individuals with age $a$, longevity $x$ at time $t$, and model the age and mortality dynamics via the McKendrick von-Foerster model

\begin{equation}
\label{PDE}
\displaystyle{\frac{\partial u}{\partial t} + \frac{\partial u}{\partial a} = -\mu(a,x) u}.
\end{equation}
\\
The mating and mutation dynamics are addressed in the boundary condition, given by

\begin{equation}
\label{BC}
u(0,x,t) = \frac{1}{\xoverline{S}_f(t) \xoverline{S}_m(t)} \int_{-\infty}^\infty S_f(y,t) \int_{-\infty}^\infty S_m(\xoverline{z},t) N \left(\frac{\xoverline{z}+y}{2},x,\varepsilon^2 \right) \, d\xoverline{z} \, dy.
\end{equation}
\\
{The boundary condition gives the total density of offspring with longevity $x$ entering the system at any time $t$. We give an explanation of its form below.} The functions $S_f(x)$ and $S_m(x)$ respectively denote the density of births by fertile females with longevity $x$ and paternities by males with longevity $x$, that is,

\begin{equation}
\begin{aligned}
S_f (x,t) &= \int_{\tau_f}^{\hat{\tau}_f} b(a,x) u(a,x,t) \, da,\\
S_m (x,t) &= \int_{\tau_m}^{\hat{\tau}_m} \phi(a,x) u(a,x,t) \, da.
\end{aligned}
\end{equation}
\\
The functions $\xoverline{S}_f(t):= \int_{-\infty}^{\infty} S_f \, dx$ and $\xoverline{S}_m(t):= \int_{-\infty}^{\infty} S_m \, dx$ represent the total density of births and a measure of total male competitiveness at time $t$ respectively. The last factor in the integrand $N(x,\mu,\varepsilon^2)$ denotes the normal density function with mean $\mu$ and variance $\varepsilon^2$; {this factor governs the mating and mutation dynamics, where the parameter $\varepsilon$ in Eq.~(\ref{BC}) denotes the mutation rate, which we assume to be small. The expression ${S_m (\xoverline{z},t)}/{\xoverline{S}_m (t)}$ represents the probability of a female mating with a male with longevity $\xoverline{z}$, or equivalently, the probability of a male with longevity $\xoverline{z}$ securing a mate. The total paternities by males with longevity $\xoverline{z}$ is then given by $S_m(\xoverline{z},t) S_f(y,t)/ \xoverline{S}_m(t)$. Including the mutation factor $N((\xoverline{z}+y)/2,x,\varepsilon^2)$ and integrating over $y$ gives the total density of offspring with longevity $x$ entering the system at any time. Finally, we divide by $\xoverline{S}_f(t)$ to regulate the number of offspring entering the system, which serves to keep the total population density finite as described in Section \ref{setup}. This process results in the boundary condition Eq.~(\ref{BC}).}
\\
\\
To simplify Eq.~(\ref{BC}), we let $\xoverline{z} = 2z-y$ to obtain

\begin{equation}
\label{BC2}
\begin{aligned}
u(0,x) &= 2 \frac{1}{\xoverline{S}_f \xoverline{S}_m} \int_{-\infty}^\infty S_f(y) \int_{-\infty}^\infty S_m(2z-y) N \left(z,x,\varepsilon \right) \, dz \, dy,\\
& = \frac{2}{\xoverline{S}_f \xoverline{S}_m} S_f(x) \ast S_m(x) [2x] \ast N(x,0,\varepsilon^2)[x],
\end{aligned}
\end{equation}
\\
where $\ast$ denotes the convolution operator, i.e. $f(x) \ast g(x) [x] = \int_{-\infty}^\infty f(y)g(x-y) \, dy$.
\\
\\
Since the mortality and ageing dynamics in Eq.~(\ref{PDE}) and the evolutionary dynamics in Eq.~(\ref{BC2}) are on different timescales, we simplify the system by assuming that the population is always at a stable age distribution. Thus, the density of births and paternities for specific longevity values, given above by $S_f$ and $S_m$, can be simplified to

\begin{equation}
\begin{aligned}
S_f (x) &= u(0,x)\int_{\tau_f}^{\hat{\tau}_f} b(a,x) \exp \left( -\int_0^a \mu(s,x) \, ds \right) \, da,\\
S_m (x) &= u(0,x)\int_{\tau_m}^{\hat{\tau}_m} \phi(a,x) \exp \left( -\int_0^a \mu(s,x) \, ds \right) \, da.
\end{aligned}
\end{equation}
\\
For convenience, we define

\begin{equation}
\label{imp}
\begin{aligned}
F(x) = \int_{\tau_f}^{\hat{\tau}_f} b(a,x) \exp \left( -\int_0^a \mu(s,x) \, ds \right) \, da,\\
M(x) = \int_{\tau_m}^{\hat{\tau}_m} \phi(a,x) \exp \left( -\int_0^a \mu(s,x) \, ds \right) \, da.
\end{aligned}
\end{equation}
\\
\\
The evolutionary dynamics of the system, given in Eq.~(\ref{BC2}), is then simplified to the following integrodifference equation

\begin{equation}
\label{convint}
\begin{aligned}
u_{n+1}(x) = 2 \left( \frac{F(x)u_n(x)}{\langle F(x),u_n(x) \rangle} \right) \ast \left( \frac{M(x)u_n(x)}{\langle M(x),u_n(x) \rangle} \right) [2x] \ast N(x,0,\varepsilon^2) [x],
\end{aligned}
\end{equation}
\\
where the angled brackets denote the $L^2$ inner product $\langle f(x),g(x) \rangle := \int_{-\infty}^\infty f(x)g(x) \, dx$. {Since $\frac{F(x)u_n(x)}{\langle F(x),u_n(x) \rangle}$ and $\frac{M(x)u_n(x)}{\langle M(x),u_n(x) \rangle}$ both integrate to 1, is it instructive to view them as probability density functions. The term $2 \left( \frac{F(x)u_n(x)}{\langle F(x),u_n(x) \rangle} \right) \ast \left( \frac{M(x)u_n(x)}{\langle M(x),u_n(x) \rangle} \right) [2x]$ can then be interpreted as another probability density function which lies ``in-between" $\frac{F(x)u_n(x)}{\langle F(x),u_n(x) \rangle}$ and $\frac{M(x)u_n(x)}{\langle M(x),u_n(x) \rangle}$ assuming $u_n(x)$ is bell-shaped, see Figure \ref{fig1}}. Convolving with $N(x,0,\varepsilon^2) [x]$, the term responsible for mutations, ``smooths" this function. {The convolution of probability density functions results in another probability density function, thus} $u_{n+1}(x)$ integrates to 1 and can be viewed as a probability density function for all $n$. We provide an illustration of one hypothetical iteration of Eq.~(\ref{convint}) in Figures \ref{fig1}-\ref{fig2}, {where one iteration encapsulates the mating, mutation and birth dynamics for one generation. The process illustrated in Figures \ref{fig1}-\ref{fig2} is a continuous perspective on the discrete problem described in Section \ref{setup}. Its main advantage is that the computation of convolutions of functions is much faster and more elegant than the agent-based approach of simulating discrete individuals. We note that this approach does take into account the biases in the mating sex ratio, even though there is no age-structure in Eq.~(\ref{convint}).}
\\
\\
The function $F(x)$ gives the expected number of births by a female with longevity $x$ in her lifetime, while $M(x)$ {gives a value proportional to the fraction of available mating opportunities a male with longevity $x$ will have in his lifetime. We let $F(x)$ and $M(x)$ be measures of fitness for females and males respectively, although we note that $M(x)$ is technically not a measure of reproductive success, instead it is a measure of the expected proportion of matings a male with longevity $x$ will have in his lifetime}. {We note that a special case of $F(x)$ and $M(x)$ is when the tradeoff functions are independent of age $a$, in which case $F(x) = b(x) k_1(x)$ and $M(x) = \phi(x) k_2(x)$, where $k_1(x)$ and $k_2(x)$ are the expected number of years survived in the fertile ages for females and males respectively with longevity trait value $x$.}
\\
\\
A natural question is, to which longevity value $x$ does the population converge for given parameter values and tradeoff functions of the model? {Through the stability analysis in Section \ref{stability}, we show that the answer} is simply the longevity value $x$ that maximises the product $F(x)M(x)$. In fact, $F(x)M(x)$ represents the fitness landscape of the scenario outlined in Section \ref{setup}, and so one can examine the shape of $F(x)M(x)$ and infer the optimal longevity value $x$ of the population without performing a simulation of the PDE. {The magnitude of the sexual conflict in the system can be inferred from a comparison of $F(x)$, $M(x)$ and $F(x)M(x)$, that is, a comparison of the female, male and combined fitness landscapes}. Furthermore, we show that the speed of convergence to the optimum of the fitness landscape is proportional to the slope of the fitness landscape $F(x)M(x)$. We generalise these results for an arbitrary number of traits in Section \ref{ntrait}.

%\begin{figure}[H]
%  \centerline{
%  \includegraphics[width=0.7\textwidth]{fig1.pdf}}
%  \caption{Plot illustrating the process described in Eq.~(\ref{convint}) {for hypothetical functions $F(x)$ and $M(x)$. We note that the height of the individual curves is irrelevant; the important aspect is that they integrate to 1, since they describe distributions across trait-space.}}
%  \label{fig1}
%\end{figure}
%
%\begin{figure}[H]
%  \centerline{
%  \includegraphics[width=0.7\textwidth]{fig2.pdf}}
%  \caption{Plot illustrating the process described in Eq.~(\ref{convint}). The black curve, representing $u_{n+1}(x)$, is obtained by convolving the function given in red with a normal density function with zero mean and finite variance.}
%  \label{fig2}
%\end{figure}

\section{Including grandmothering}
\label{GMsect}

\begin{figure}[H]
  \centerline{
  \subfloat[]{\label{GMprop}\includegraphics[width=0.55\textwidth]{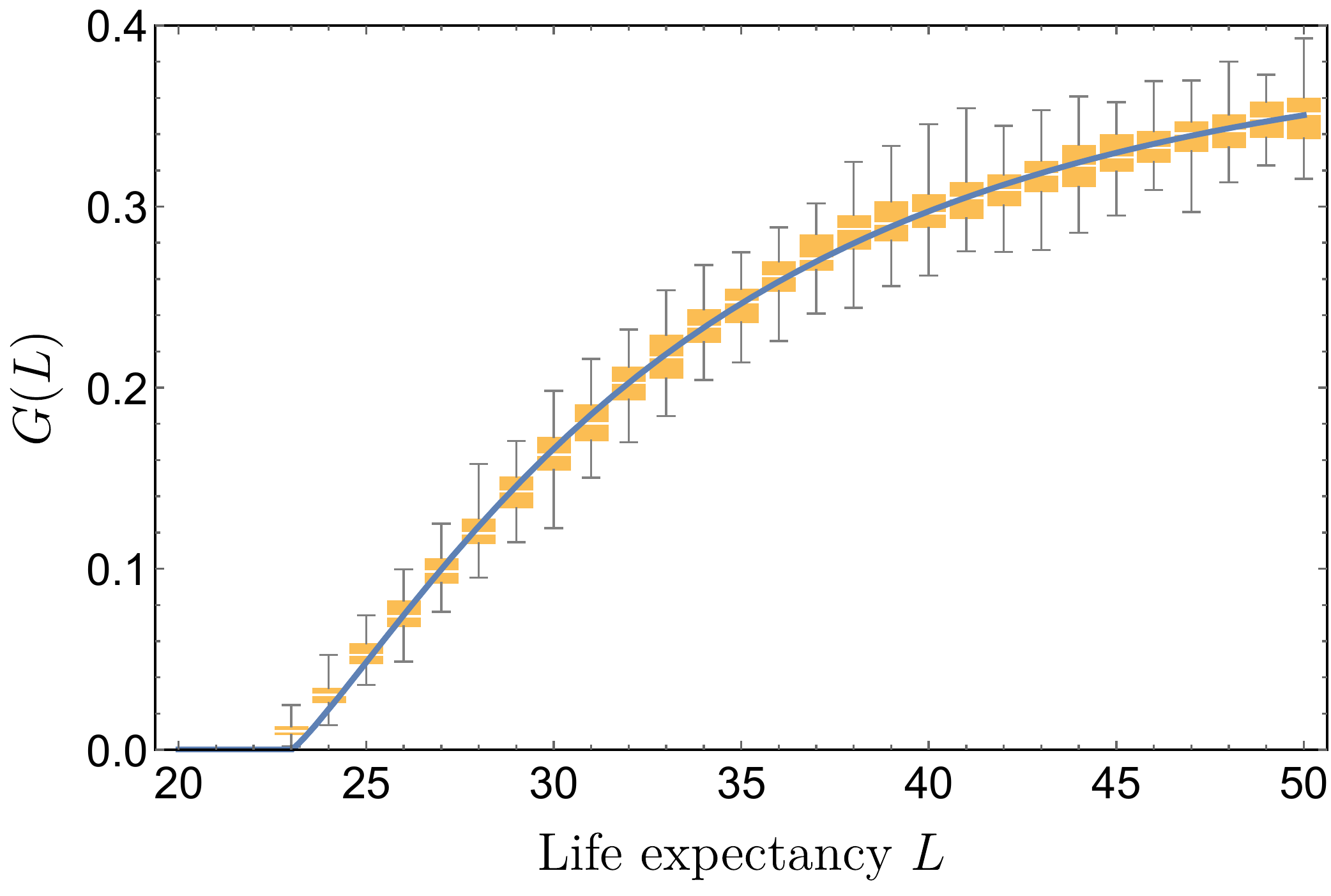}}
  \hspace{2mm}
  \subfloat[]{\label{GMfitness}\includegraphics[width=0.55\textwidth]{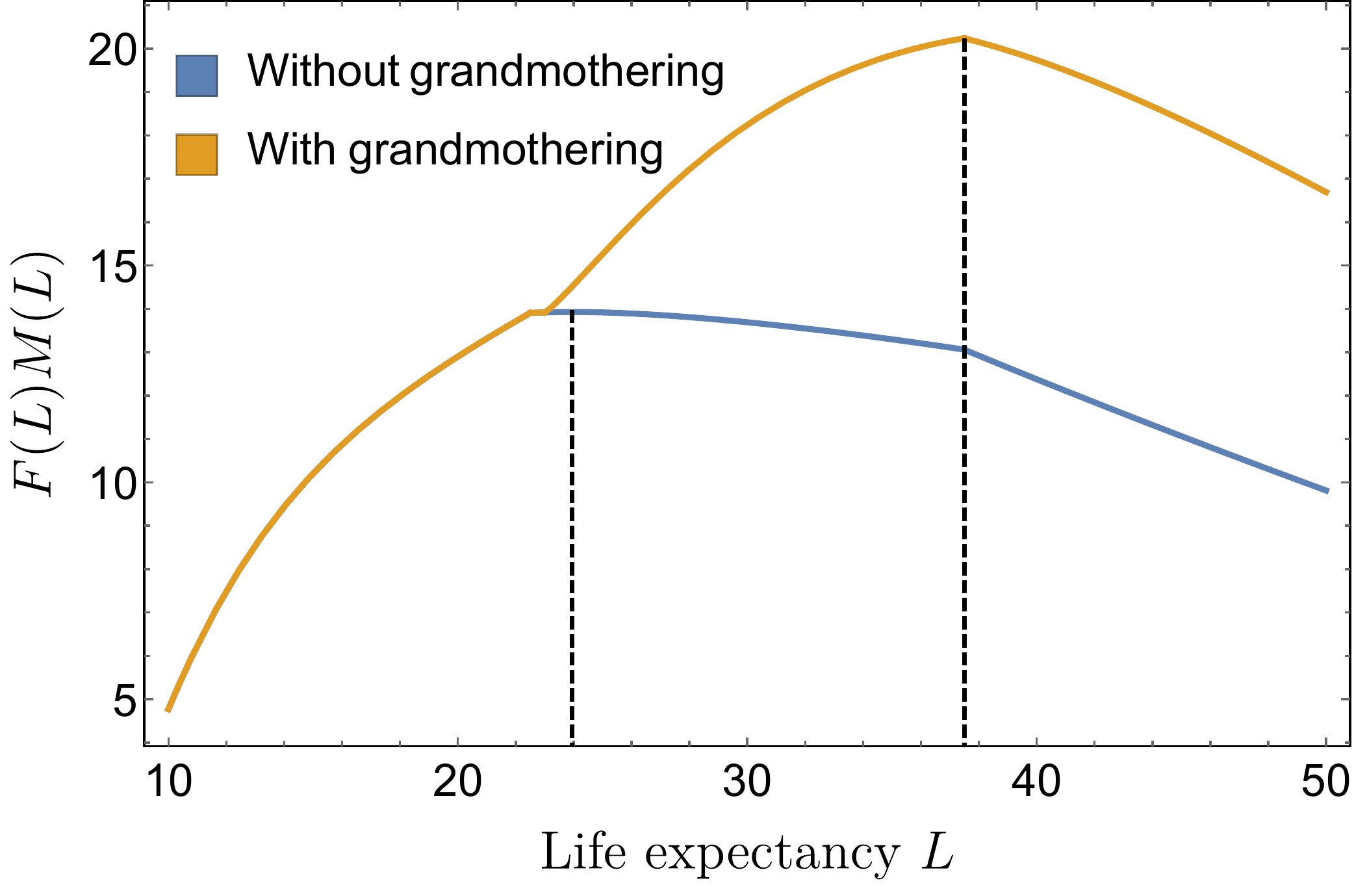}}}
  \caption{Figure (a) shows box plots of the proportion of fertile females with a surviving post-fertile mother when the population has reached stable age distribution in the ABM. The function $G(x) = \frac{(x-23)^a}{b+(x-23)^c}$ is used as a model for $G(x)$, where $a = 1.164$, $b=43.83$ and $c=1.36$. Figure (b) shows fitness landscapes of the population with and without grandmothering. The dashed lines correspond to the maxima of each curve. When there is no grandmothering, the population converges to a life expectancy of approximately 24, whereas the population transitions to a life expectancy of approximately 38 with grandmothering.}
  \label{fig3}
\end{figure}

In this section, we consider the same problem as in Section \ref{setup}, except that we allow for grandmothering, i.e. provisioning from post-fertile females. We let $L$, an individual's life expectancy, be the only heritable trait in the population. Similar to Chan et al. \cite{170}, we assume that post-fertile females take care of all their matrilineal grandchildren and that grandmothers boost their daughters' birth rates. The birth rate $b(a,L)$ in Eq.~(\ref{imp}) is then of the form $B(G(L)) \bar{b}(a,L)$, where $\bar{b}(a,L)$ is the base birth rate without the help of grandmothers, $G(L)$ is the proportion of fertile females covered by a post-fertile mother for a fixed longevity $L$, assuming a stable age distribution, and $B(x) = mx + (1-x)$ represents the boost in fertility. Following Chan et al. \cite{170}, we let the base birth rate $\bar{b}(L)$ be independent of age and require it to be equal to 0.3/year at $L=14$ and 0.11/year at $L=34$. {Also guided by Chan et al. \cite{170},} we set $m=3$, so that grandmothers boost their daughters' birth rates by a factor of 3, thus increasing the birth rate to 0.33/year for an individual with a life expectancy of 34 years, who is also supported by a post-fertile mother.
\\
\\
We use the parameter settings of Kim et al. \cite{117} for $\tau_f(L)$, $\tau_m$, $\hat{\tau}_f$, $\hat{\tau}_m(L)$ and $\mu(L)$. These parameters are listed in Table \ref{lifehistoryparatable}. {We include the assumption from Kim et al. \cite{117} that individuals are frail, and thus exit the population, at the age $\tau_T = \min(2L,75)$ to prevent individuals living to unrealistic ages. Moreover, males are fertile from sexual maturity $\tau_m$ until frailty $\tau_T$.} We note that the monotonically increasing nature of the age of female sexual maturity $\tau_f(L)$ is based on Charnov's model of mammalian life history, which predicts an increase in maturation age with greater longevities \cite{129}. The male age of sexual maturity is kept constant for simplicity.
\\
\\
{Since} we could not find a straightforward analytic way to calculate $G(L)$ given arbitrary parameter values, we approximate $G(L)$ by using an agent-based model (ABM). The ABM runs as follows: First a longevity value $L$ is chosen and is fixed for the entire simulation. {All individuals have the same longevity $L$ and there is zero probability for mutations.} Females are fertile from $\tau_f(L)$ to $\hat{\tau}_f$. {Due to longevity being fixed and the assumption in Section \ref{setup} that every female is guaranteed to mate in every time step, there is no need for the ABM to track males; tracking males is only essential if $L$ is allowed to evolve}. At each time step, every fertile female gives birth to one offspring with probability $1-\exp(-\bar{b}(x))$ if she does not have a surviving post-fertile mother or with probability $1-\exp(-3\bar{b}(x))$ if she does. If there are over 200 newborns, then newborns are randomly removed from the population until only 200 remain. Individuals die with probability $1-\exp(-\mu(x))$ and any individuals past the age of $\min(2L,75)$ are removed from the population. Figure \ref{GMprop} shows 100 samples of $G(x)$ for each integer $L$ past 22, since there are no grandmothers unless $L \geq 23$ due to {$\hat{\tau}_f = \min(2L,45)$}. We fit the function

\begin{equation}
G(x) = 
\begin{cases}
\begin{aligned}
\frac{(x-23)^a}{b+(x-23)^c} \quad\quad &\text{if } x \geq 23,\\
0 \quad\quad &\text{if } x<23,
\end{aligned}
\end{cases}
\end{equation}
\\
through the means of each data set collected in Figure \ref{GMprop} and use the method of least squares to obtain $a = 1.164$, $b=43.83$ and $c=1.36$.
\\
\\
Following Kim et al. \cite{116}, {we let the male fertility-longevity tradeoff $\phi(L)$, the relative probability for a male with longevity $L$ to secure a mate, to decrease exponentially with longevity by requiring it to satisfy} $\phi'(L) = \delta(L)\phi(L)$. The function $\phi(L)$ is then uniquely defined, up to a scaling factor, by

\begin{equation}
\label{maletrade}
\phi(L) = \exp \left( \int_{L_0}^L \delta (s) \, ds \right),
\end{equation}
\\
where we choose $L_0=20$ and $\delta(L) = -0.4\exp(-0.087 L)$.

\begin{table}[H]
  \centering
  \caption{Parameter values}
    \begin{tabular}{lll}
    \toprule
    Symbol & Definition & Value\\
    \midrule
    $m$  & Benefit received by fertile female from grandmothering & $3$  \\
    $\tau_T(L)$  & Age of frailty & $\min(2L,75)$  \\
    $\tau_f(L)$	& Age of female sexual maturity & $L/2.5 + 2$ \\
    $\tau_m$	& Age of male sexual maturity & 15 \\
    $\hat{\tau}_f$	& Age female fertility ends & 45 \\
    $\hat{\tau}_m(L)$ & Age male fertility ends & $\min(2L,75)$ (equal to $\tau_T$) \\
    $\phi(L)$ & Male fertility-longevity tradeoff & Function of $L$, see Eq.~(\ref{maletrade}) \\
    $\mu(L)$ & Mortality rate & $1/L$ \\
    $\bar{b}(L)$ & Birth rate & $4.522/L - 0.023$ \\
    \bottomrule
    \end{tabular}
  \label{lifehistoryparatable}
    \\[5pt]
\end{table}

\begin{figure}[H]
  \centerline{
  \subfloat[]{\label{GMfitness2}\includegraphics[width=0.55\textwidth]{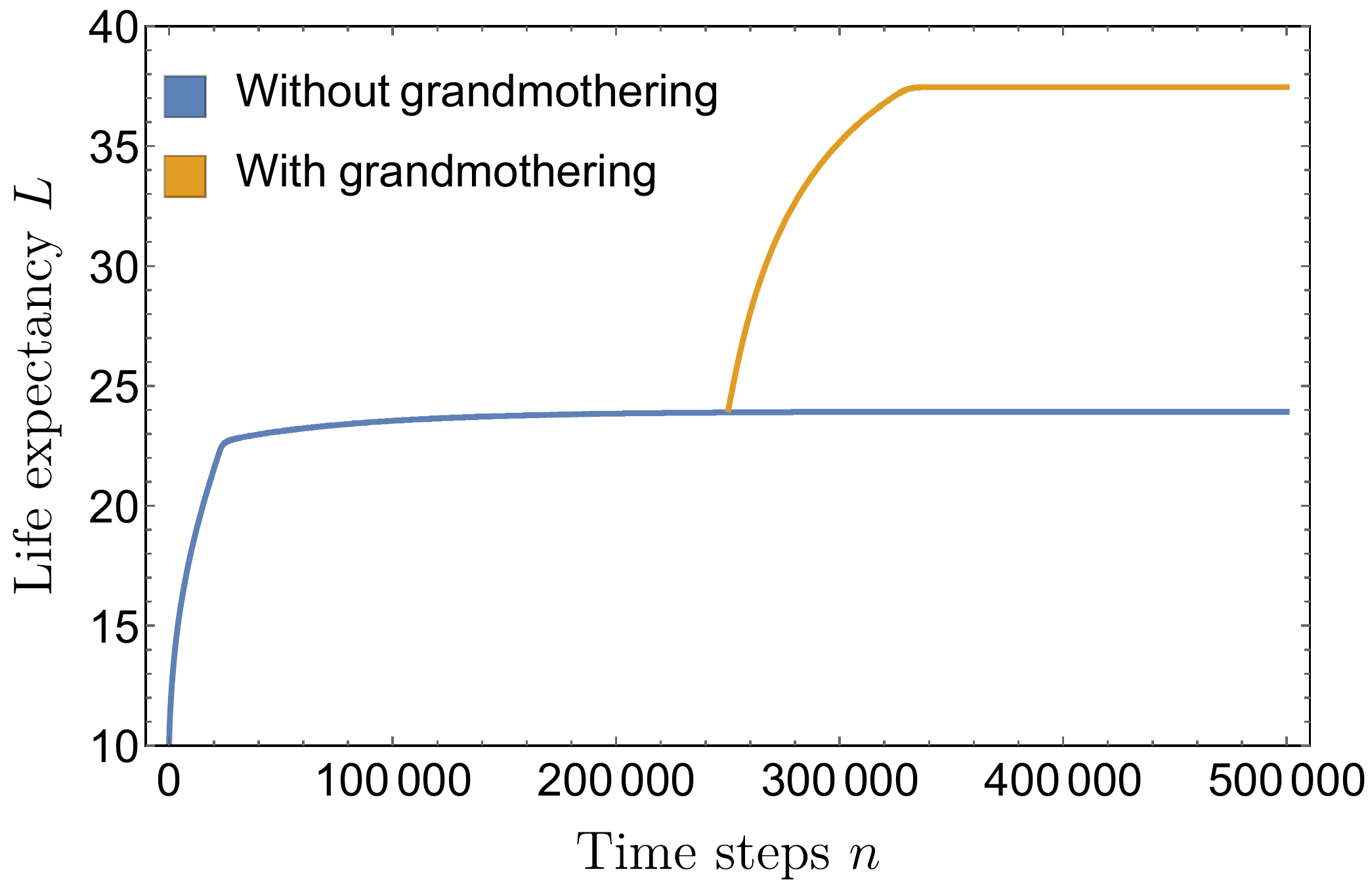}}
  \hspace{2mm}
  \subfloat[]{\label{GMfitness3}\includegraphics[width=0.505\textwidth]{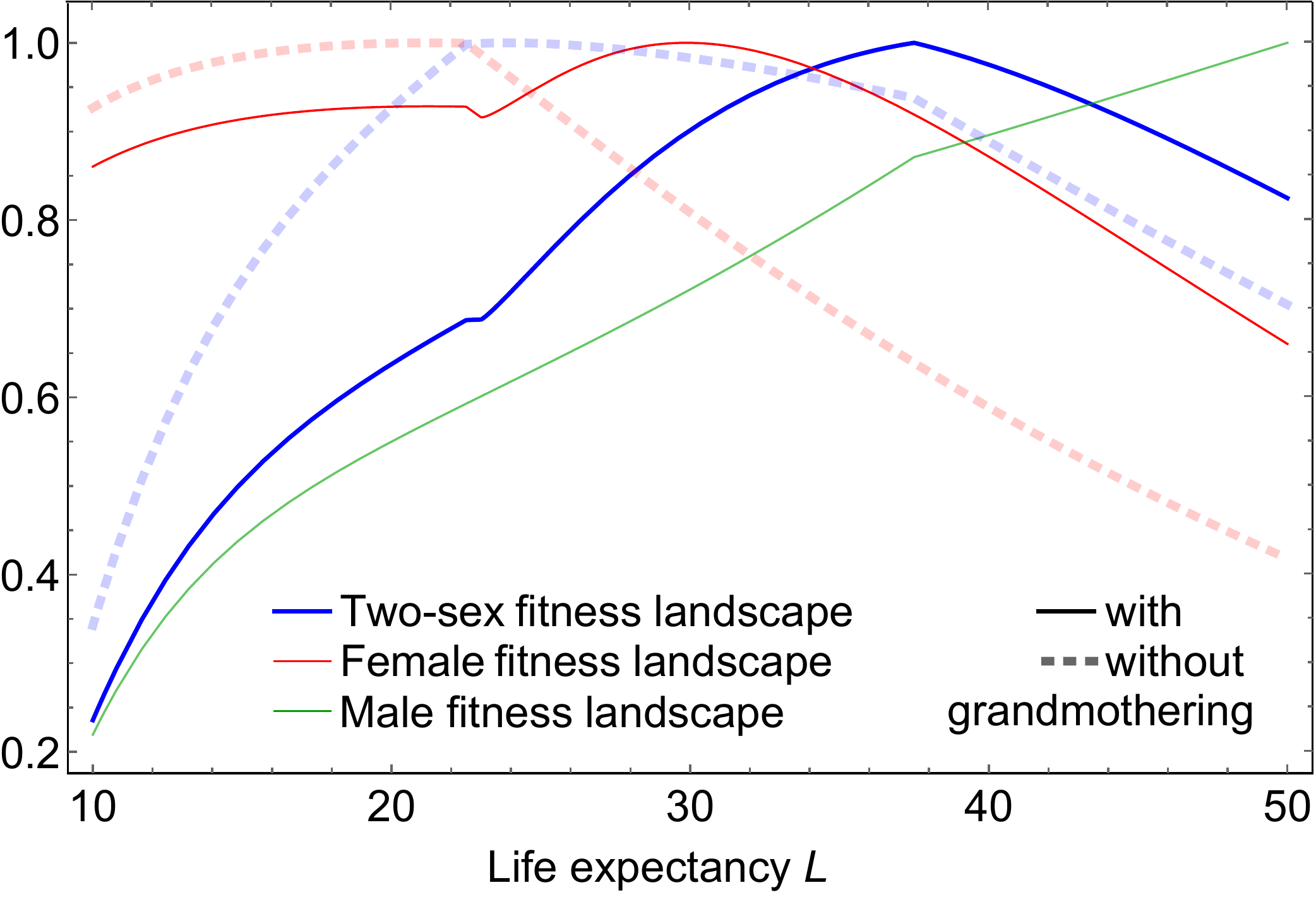}}}
  \caption{Figure (a) shows plots of the first moment of $u_n(L)$, as defined in Eq.~(\ref{convint}), with and without grandmothering, with $\varepsilon^2=0.025$. When the population without grandmothering (blue curve) reached equilibrium, grandmothering was allowed for (orange curve). The equilibria match the maxima of the fitness landscapes shown in Figure \ref{GMfitness}, as expected from the proof in Section \ref{stability}. Figure (b) shows plots of the female fitness landscape $F(L)$, male fitness landscape $M(L)$ and the two-sex fitness landscape $F(L)M(L)${, with grandmothering (solid) and without grandmothering (dashed).} Each curve is scaled such that the maximum value is 1. This provides a visualisation of the magnitude of the sexual conflict in the system.}
  \label{fig4}
\end{figure}

\noindent To examine the life expectancy {that is selected in the population} with and without grandmothering, we compute $F(L)M(L)$. As shown in Figure \ref{GMfitness}, the fitness landscape is the same before life expectancies of $L=23${, since $L=23$ is the earliest life expectancy at which the population has any post-fertile females.} However, for life expectancies greater than 23, grandmothering alters the fitness landscape and enables a transition to a unique optimum corresponding to a higher longevity.
\\
\\
We confirm that the population converges to the maxima of the fitness landscape in Figure \ref{GMfitness2}, where we show the expected trait value of the population over time, calculated via Eq.~(\ref{convint}). Moreover, to elucidate the role of the sexual conflict in the system, Figure \ref{GMfitness3} provides a way to visualise the magnitude of the sexual conflict by comparing the female fitness function $F(L)$, male fitness function $M(L)$ and the two-sex fitness landscape $F(L)M(L)$. From Figure \ref{GMfitness3}, male fitness increases concurrently with longevity $L$; however, this is counterbalanced by a decrease in female fitness, which creates the unique maximum point in the combined two-sex fitness landscape. One can observe the optima preferred by the males and females through their respective fitness landscapes, and also the compromise adopted by the population overall that lies between these two optima, as shown by the two-sex fitness landscape. We note that the male fitness landscape is the same regardless of whether there is grandmothering in the population, since we have made the assumption that the only function of grandmothers in the population is to boost their daughter's birth rate.

\section{Discussion}

We have presented a computationally inexpensive method to compute the fitness landscape of the problem presented in Section \ref{setup} and have applied it to determine if grandmothering in a population can enable an evolutionary trajectory to higher longevities. Previous studies of this problem have included both short- and long-time scale components, which result in computationally expensive simulations and hinders inference from the model. Our method essentially compacts the short-time scale components of the problem, such as mortality and ageing, by assuming a stable age-distribution, while focusing on the evolution of traits in the population, which we assume to be of a much longer-time scale. The resultant equation describing the evolutionary dynamics, given by Eq.~(\ref{convint}), can be viewed as a dynamical systems generalisation of Price's Equation that includes two-sexes and multiple traits.
\\
\\
{A difficulty encountered by previous studies in the literature is assessing the role of the sexual conflict in the problem. This is because the sexual conflict is a manifestation of the many different components; specifically, the mating sex ratio (which is itself determined by the male and female fertile ages, and the mortality rate) and the male and female tradeoffs with longevity. {Although it is tempting to analyse the sexual conflict through the mating sex ratio, it is not the correct approach; we stress that since a given mating sex ratio value can be achieved through a variety of different fertile ages and mortality rates, there is not a unique correspondence between mating sex ratio values and fitness landscapes.} This is evident from the form of the female fitness landscape $F(L)$ and male fitness landscape $M(L)$ (see Eq.~(\ref{imp})). {More specifically, the function $k(L)$, denoting the mating sex ratio at longevity $L$ and given by

\begin{equation}
k(L) = \left(\int_{\tau_m}^{\hat{\tau}_m} \exp \left( -\int_0^a \mu(s,L) \, ds \right) \, da \right) \Bigg/ \left( \int_{\tau_f}^{\hat{\tau}_f} \exp \left( -\int_0^a \mu(s,L) \, ds \right) \, da \right),
\end{equation}
\\
is not guaranteed to be invertible. This implies that the same mating sex ratio value may be obtainable at different longevity $L$ values.} Consequently, although the sexual conflict is clearly present in the system, it is not an aspect that can be accurately represented through a single indicator and this hinders its analysis. In this study, we provide the next best approach: a method to compute the female and male fitness landscapes separately, and also the combined fitness landscape. Through this, one can observe how each sex contributes to the resulting evolutionary dynamics of the overall problem and how the male and female fitness landscapes vary with the mating sex ratio. In particular, when the female and male tradeoffs are quantitatively known, our method allows one to assess how much contribution a change in mating sex ratio contributes to the fitness landscape of a population.}
\\
\\
A limiting assumption of our model is that fertile males compete over every fertile female for a chance at paternity regardless of the mating sex ratio. However, a shifting mating sex ratio will influence the availability of partners due to the asymmetry in the male and female fertile ages. Hence, mating strategies are likely to change along with the mating sex ratio (see Schacht \& Bell \cite{172}, {Coxworth et al. \cite{169} and Loo et al. \cite{181} for examples}). Thus, an avenue for future work is including such changes in mating strategies in response to a shifting mating sex ratio via the evolution of longevity in a population. This could in turn have important consequences on the evolutionary trajectory the population follows towards higher longevities.

\begin{acknowledgements}
MHC and PSK were supported by the Australian Research Council, Discovery Project (DP160101597).
\end{acknowledgements}

\bibliographystyle{apalike}
\bibliography{references}

\section{Appendix}

\subsection{Stability}
\label{stability}

We are interested in the equilibria of the system and how {they are} affected by the shape of $F(x)$ and $M(x)$. Specifically, we wish to find the expected value of $u_n (x)$ as $n \to \infty$. To examine this, we consider the limiting case $\varepsilon \to 0$, which represents a scenario where the probability for mutations in $x$ is zero. This yields the system

\begin{equation}
\label{convintnomut}
u_{n+1}(x) = 2 \left( \frac{F(x)u_n(x)}{\langle F(x),u_n(x) \rangle} \right) \ast \left( \frac{M(x)u_n(x)}{\langle M(x),u_n(x) \rangle} \right) [2x].
\end{equation}
\\
Since Eq.~(\ref{convint}) is a regular perturbation problem with respect to $\varepsilon$, we expect that Eq.~(\ref{convintnomut}) is an accurate approximation of Eq.~(\ref{convint}). One can check that this has an infinite number of fixed points in the form $u^*(x) = \delta(x-\mu)$, for any constant $\mu$. However, assuming that the initial condition $u_0(x)$ is strictly positive, bounded and integrable, the sequence of functions $u_n(x)$ only converges to $u^*(x)$ for particular values of $\mu$. To determine such values of $\mu$, we compute the first moment of both sides of Eq.~(\ref{convintnomut}), which we denote by $\Lambda$. It is well known that if $X$ and $Y$ have the density functions $f$ and $g$ respectively, then the density function for $\frac{1}{2}(X+Y)$ is $2 (f(x)\ast g(x))[2x]$. We have that $\Lambda (2 (f(x)\ast g(x))[2x]) = \text{\bf{E}} (\frac{1}{2} (X+Y)) = \frac{1}{2}(\Lambda f + \Lambda g)$. Thus,

\begin{equation}
\label{Erecur}
\Lambda(u_{n+1}(x)) = \frac{1}{2} \left( \Lambda \left( \frac{F(x)u_n(x)}{\langle F(x),u_n(x) \rangle} \right) + \Lambda \left( \frac{M(x)u_n(x)}{\langle M(x),u_n(x) \rangle} \right) \right).
\end{equation}
\\
Due to the fixed points being in the form of $u^*(x) = \delta(x-\mu)$, for any constant $\mu$, to assess stability we fix $u_n(x)$ in the following box function form

\begin{equation}
u_n(x;\mu_n,\alpha) =
\begin{cases}
\frac{1}{2\alpha} \qquad \text{if} \qquad x \in [\mu_n-\alpha, \mu_n+\alpha],\\
0 \qquad\text{otherwise}.
\end{cases}
\end{equation}
\\
where $\alpha>0$ is a small parameter and $\mu_n$ is the expected value of $u_n(x;\mu,\alpha)$ by construction. This acts as a perturbation of $\delta(x-\mu)$ for $\alpha$ small. Substituting this into Eq.~(\ref{Erecur}), we obtain 

\begin{equation}
\label{expectworking}
\mu_{n+1} = \frac{1}{2} \left( \frac{\int_{\mu_n-\alpha}^{\mu_n+\alpha} x F(x) \, dx}{\int_{\mu_n-\alpha}^{\mu_n+\alpha} F(x) \, dx} + \frac{\int_{\mu_n-\alpha}^{\mu_n+\alpha} x M(x) \, dx}{\int_{\mu_n-\alpha}^{\mu_n+\alpha} M(x) \, dx} \right).
\end{equation}
\\
We require that $\mu_n$ such that it is a fixed point of Eq.~(\ref{Erecur}). Since $\alpha$ is arbitrarily small, we take the first order approximation of $F(x)$ and $M(x)$ about $\mu_n$, that is, we let $F(x) = F(\mu_n) + F'(\mu_n)(x-\mu_n)$ and $M(x) = M(\mu_n) + M'(\mu_n)(x-\mu_n)$. After some algebraic manipulation, Eq.~(\ref{expectworking}) simplifies to the recurrence relation

\begin{equation}
\label{expectedvaluerecur}
\mu_{n+1} = \Phi(\mu_n) := \mu_n + \frac{\alpha^2}{6} \left( \frac{\frac{d}{dx}(F(x)M(x))}{F(x)M(x)} \right) \Bigg|_{x=\mu_n}.
\end{equation}
\\
Thus, fixed points $\mu^*$ of Eq.~(\ref{expectedvaluerecur}) satisfy the condition

\begin{equation}
\label{fixedpointcondition}
\frac{d}{dx}(F(x)M(x))\Big|_{x=\mu^*} = 0.
\end{equation}
\\
To determine the stability of $\mu^*$ satisfying the above condition, we examine $\Phi(\mu^*)$. We require $|\Phi'(\mu^*)| < 1$ for local stability (see Theorem 1.5 in \cite{171}). We have that

\begin{equation}
\frac{d}{dx} \Phi(\mu^*) = 1 - \frac{\alpha^2}{6} \left( \frac{\frac{d}{dx}(F(x)M(x))}{(F(x)M(x))^2} \right) \Bigg|_{x=\mu^*}+\frac{\alpha^2}{6} \left( \frac{\frac{d^2}{dx^2}(F(x)M(x))}{F(x)M(x)} \right) \Bigg|_{x=\mu^*}.
\end{equation}
\\
Thus, fixed points satisfying Eq.~(\ref{fixedpointcondition}) are locally stable when

\begin{equation}
\label{stabilitycondition}
\frac{-12 F(x) M(x)}{\alpha^2} < \frac{d^2}{dx^2}(F(x)M(x))\Big|_{x=\mu^*} < 0.
\end{equation}
\\
We note that $\alpha$ can be chosen such that $-12/\alpha^2$ is arbitrarily negative. The conditions given in Eq.~(\ref{fixedpointcondition}) and Eq.~(\ref{stabilitycondition}) indicate that $F(x)M(x)$ represents the total fitness landscape. As $n$ increases, the first moment of $u_n(x)$ moves in the direction of increasing gradient in $F(x)M(x)$. We note that it is possible for the solution to be trapped in a local optima, i.e. the population converges to the nearest peak on the fitness landscape.

\subsection{Generalisation to $n$ traits}
\label{ntrait}

In this section we generalise the results in Section \ref{stability} to $n$ traits, where we emulate the derivation, except in $n$-dimensions. Let $\mathbf{x} = (x_1,x_2,\dots,x_n)'$ be the vector of $n$ traits. The PDE in Eq.~(\ref{PDE}) remains unchanged, except that the functions $u$ and $\mu$ map from $\mathbb{R}^n$ to $\mathbb{R}$. The boundary condition given in Eq.~(\ref{BC}) becomes

\begin{equation}
u(0,\mathbf{x}) = \frac{1}{\xoverline{S}_f \xoverline{S}_m} \int_{-\infty}^\infty \int_{-\infty}^\infty \dots \int_{-\infty}^\infty S_f(\mathbf{y}) S_m(\xoverline{\mathbf{z}}) N \left( \frac{\mathbf{y}+\xoverline{\mathbf{z}}}{2},\mathbf{x},\varepsilon^2 \right) \, d\mathbf{y} \, d\xoverline{\mathbf{z}},
\end{equation}
\\
where $b(\mathbf{x})$, $\phi({\mathbf{x}})$, $\tau_f({\mathbf{x}})$, $\tau_m({\mathbf{x}})$, $\hat{\tau}_f({\mathbf{x}})$, $\hat{\tau}_m({\mathbf{x}})$, $S_f(\mathbf{x})$ and $S_m({\mathbf{x}})$ are now functions which map from $\mathbb{R}^n$ to $\mathbb{R}$. The vectors $\mathbf{y} := (y_1,y_2,\dots,y_n)'$ and $\xoverline{\mathbf{z}} := (\xoverline{z}_1,\xoverline{z}_2,\dots,\xoverline{z}_n)'$ are integration dummy variables. Furthermore, $N\left( \mathbf{x},\mathbf{y},\varepsilon^2 \right)$ is the $n$ dimensional normal density function given by

\begin{equation}
N(\mathbf{x},\mathbf{y},\varepsilon^2) = (2\pi \varepsilon^2)^{-n/2} \exp \left( \frac{-(\mathbf{x}-\mathbf{y})^T(\mathbf{x}-\mathbf{y})}{2 \varepsilon^2} \right).
\end{equation}
\\
Lastly, $\xoverline{S}_f$ denotes $\int_{\mathbb{R}^n} S_f(\mathbf{x}) \, d\mathbf{x}$ and similarly for $\xoverline{S}_m$.
\\
\\
Letting $\xoverline{\mathbf{z}} = 2 \mathbf{z} - \mathbf{y}$, we obtain

\begin{equation}
\begin{aligned}
u(0,\mathbf{x}) &= \frac{2^n}{\xoverline{S}_f \xoverline{S}_m} \int_{-\infty}^\infty \int_{-\infty}^\infty \dots \int_{-\infty}^\infty S_f(\mathbf{y}) S_m(2{\mathbf{z}}-\mathbf{y}) N \left( \mathbf{z},\mathbf{x},\varepsilon^2 \right) \, d\mathbf{y} \, d\mathbf{z},\\
&= \frac{2^n}{\xoverline{S}_f \xoverline{S}_m} S_f(\mathbf{x}) \otimes S_m(\mathbf{x})[2\mathbf{x}] \otimes N(\mathbf{x},\mathbf{0},\varepsilon^2)[\mathbf{x}],
\end{aligned}
\end{equation}
\\
where $\otimes$ denotes the $n$-dimensional convolution operator, that is,

\begin{equation}
f(\mathbf{x}) \otimes g(\mathbf{x})[\mathbf{x}] = \int_{-\infty}^\infty \int_{-\infty}^\infty \dots \int_{-\infty}^\infty f(y_1,y_2,\dots,y_n)g(x_1 - y_1, x_2 - y_2,\dots,x_n-y_n) \, d\mathbf{y}.
\end{equation}
\\
\\
Letting $\varepsilon \to 0$ and assuming that the population is at stable age distribution for all time, that is,

\begin{equation}
\begin{aligned}
S_f(\mathbf{x},t) &= \frac{u(0,\mathbf{x},t)}{2} \int_{\tau_f}^{\hat{\tau}_f} b(a,\bm{x}) \exp \left( -\int_0^a \mu(s,\mathbf{x}) \, ds \right)\, da, \\
S_m(\mathbf{x},t) &= \frac{u(0,\mathbf{x},t)}{2} \int_{\tau_m}^{\hat{\tau}_m} \phi(a,\bm{x}) \exp \left( -\int_0^a \mu(s,\mathbf{x}) \, ds \right)\, da,
\end{aligned}
\end{equation}
\\
we obtain 

\begin{equation}
\label{nomutvec}
u_{t+1}(\mathbf{x}) = 2^n \left( \frac{F(\mathbf{x})u_t(\mathbf{x})}{\langle F,u_t \rangle} \otimes \frac{M(\mathbf{x})u_t(\mathbf{x})}{\langle M,u_t \rangle} \right)[2 \mathbf{x}],
\end{equation}
\\
where $\langle F,u_t \rangle$ denotes $\int_{\mathbb{R}^n} F(\bm{x}) u_{t}(\mathbf{x}) \, d\mathbf{x}$ and similarly for $\langle M,u_t \rangle$.
\\
\\
Computing the first moment of both sides leads to

\begin{equation}
\label{expectvec}
\Lambda(u_{t+1}(\mathbf{x})) = \frac{1}{2} \left( \Lambda \left( \frac{F(\mathbf{x})u_t(\mathbf{x})}{\langle F,u_t \rangle} \right) + \Lambda \left( \frac{M(\mathbf{x})u_t(\mathbf{x})}{\langle M,u_t \rangle} \right) \right),
\end{equation}
\\
where we have defined

\begin{equation}
\Lambda(f(\mathbf{x})) = 
\colvec{4}{\int_{\mathbb{R}^n} x_1 f(\mathbf{x}) \, d\mathbf{x}}{\int_{\mathbb{R}^n} x_2 f(\mathbf{x}) \, d\mathbf{x}}{\vdots}{\int_{\mathbb{R}^n} x_n f(\mathbf{x}) \, d\mathbf{x}}.
\end{equation}
\\
Since the fixed points of Eq.~(\ref{nomutvec}) are in the form $u^* = \delta(x_1-\mu_1) \delta(x_2-\mu_2) \dots \delta(x_3 - \mu_3)$, we examine stability by perturbing this fixed point by setting $u_{t}(\mathbf{x})$ to the following box function form

\begin{equation}
u_t(\mathbf{x},\bm{\mu}_t,\alpha) =
\begin{cases}
\begin{aligned}
&\frac{1}{(2 \alpha)^n} \qquad \text{if } \qquad x \in A,\\
&\quad0 \qquad \text{otherwise,}
\end{aligned}
\end{cases}
\end{equation}
\\
where $A$ is the $n$-dimensional box centered at coordinates given by $\bm{\mu}_t$ with length $2 \alpha$ along each dimension. Thus, the volume of $A$ is equal to 1. Substituting this into Eq.~(\ref{expectvec}), we obtain

\begin{equation}
\label{temp}
\bm{\mu}_{t+1} = \frac{1}{2} \left( \frac{\int_{\bm{\mu}_t-\alpha}^{\bm{\mu}_t+\alpha} \mathbf{x}F(\mathbf{x}) \, d\mathbf{x}}{\int_{\bm{\mu}_t-\alpha}^{\bm{\mu}_t+\alpha} F(\mathbf{x}) \, d\mathbf{x}} + \frac{\int_{\bm{\mu}_t-\alpha}^{\bm{\mu}_t+\alpha} \mathbf{x}M(\mathbf{x}) \, d\mathbf{x}}{\int_{\bm{\mu}_t-\alpha}^{\bm{\mu}_t+\alpha} M(\mathbf{x}) \, d\mathbf{x}}\right)
\end{equation}
\\
where 
\begin{equation}
\int_{\bm{\mu}_t-\alpha}^{\bm{\mu}_t+\alpha} F(\mathbf{x}) \, d\mathbf{x} = \int_{\mu^n_t-\alpha}^{\mu^n_t+\alpha}\int_{\mu^{n-1}_{t}-\alpha}^{\mu^{n-1}_{t-1}+\alpha} \dots \int_{\mu^1_1-\alpha}^{\mu^1_1+\alpha} F(\mathbf{x}) \, dx_1 \, dx_2 \, \dots \, dx_n
\end{equation}
\\
and similarly for the other integrals in Eq.~(\ref{temp}). The superscript on $\bm{\mu}_t$ denotes the index, for example, $\mu^n_t$ denotes the $n$th element of $\bm{\mu}_t$. Since $\alpha$ is arbitrarily small, we approximate $F(\mathbf{x})$ and $M(\mathbf{x})$ to first order, that is, we let $F(\mathbf{x}) = F(\bm{\mu}) + \frac{\partial}{\partial x_1} F(\mu_1)(x_1-\mu_1) + \frac{\partial}{\partial x_2} F(\mu_2)(x_2-\mu_2) + \dots + \frac{\partial}{\partial x_n} F(\mu_n)(x_n-\mu_n)$ and $M(\mathbf{x}) = M(\bm{\mu}) + \frac{\partial}{\partial x_1} M(\mu_1)(x_1-\mu_1) + \frac{\partial}{\partial x_2} M(\mu_2)(x_2-\mu_2) + \dots + \frac{\partial}{\partial x_n} M(\mu_n)(x_n-\mu_n)$. Substituting this into Eq.~(\ref{temp}) and simplifying, we obtain

\begin{equation}
\label{recurvec}
\bm{\mu}_{t+1} = \bm{\mu}_t + \frac{\alpha^2}{6 {F(\bm{\mu}_t)M(\bm{\mu}_t)}} \nabla (F(\mathbf{x})M(\mathbf{x}))\bigg|_{\mathbf{x}=\mathbf{\bm{\mu}_t}}.
\end{equation}
\\
Thus, fixed points of Eq.~(\ref{recurvec}) are also critical points of $F(\mathbf{x})M(\mathbf{x})$. In order to assess local stability, we examine the Jacobian matrix $\bm{J}$ of Eq.~(\ref{recurvec}) evaluated at the critical points $\bm{x^*}$ of $F(\mathbf{x})M(\mathbf{x})$. It is given by

\begin{equation}
\label{jac}
\bm{J} \Big | _{\mathbf{x}=\mathbf{\bm{x^*}}} = \bm{I} + \frac{\alpha^2}{6F(\mathbf{x^*})M(\mathbf{x^*})}\bm{H}(F(\mathbf{x})M(\mathbf{x}))\Big | _{\mathbf{x}=\mathbf{\bm{x^*}}},
\end{equation}
\\
where $\bm{H}(F(\mathbf{x})M(\mathbf{x}))$ denotes the Hessian matrix of $(F(\mathbf{x})M(\mathbf{x})$ and $\bm{I}$ denotes the identity matrix. For critical points to be locally stable, we require that all eigenvalues $\lambda$ of the matrix given in Eq.~({\ref{jac}}) to satisfy $|\lambda| < 1$. Since we have that

\begin{equation}
\begin{aligned}
0 &= \det \left( \bm{I} + \frac{\alpha^2}{6F(\mathbf{x^*})M(\mathbf{x^*})}\bm{H}(F(\mathbf{x})M(\mathbf{x}))\Big | _{\mathbf{x}=\mathbf{\bm{x^*}}} - \lambda \bm{I} \right),\\
&= \det \left( \frac{\alpha^2}{6F(\mathbf{x^*})M(\mathbf{x^*})}\bm{H}(F(\mathbf{x})M(\mathbf{x}))\Big | _{\mathbf{x}=\mathbf{\bm{x^*}}} - (\lambda - 1) \bm{I} \right),
\end{aligned}
\end{equation}
\\
if all eigenvalues $\lambda$ of Eq.~({\ref{jac}}) satisfy $|\lambda| < 1$, then all eigenvalues of $\bm{H}(F(\mathbf{x})M(\mathbf{x}))\Big | _{\mathbf{x}=\mathbf{\bm{x^*}}}$ are negative. Thus, locally stable fixed points of Eq.~(\ref{recurvec}) correspond to local maxima of $F(\mathbf{x})M(\mathbf{x})$. From Eq.~(\ref{recurvec}), we can see that the population converges to locally stable fixed points along the gradient of $F(\mathbf{x})M(\mathbf{x})$, i.e. the direction of steepest ascent on the fitness landscape.

\end{document}